\documentclass[seceq,supplement]{ptptex}
\title{
Are there long-time tails in granular flows?
}

\author{
Hisao \textsc{Hayakawa} and Michio \textsc{Otsuki} %
}

\inst{
Yukawa Institute for Theoretical Physics, Kyoto University,  Kitashirakawaoiwake-cho, Sakyo,
Kyoto 606-8502, JAPAN
}
\usepackage{amsmath}
\usepackage{dcolumn}
\usepackage[xdvi]{graphicx}%
\makeatletter

\def\btt#1{\texttt{\@backslashchar#1}}%
\DeclareRobustCommand\bblash{\btt{\@backslashchar}}%

\newcommand{\bv}[1]{{\boldsymbol #1}}

\abst{
The long-time behaviors of the velocity autocorrelation function (VACF)
for granular flows based on the optimal velocity fluid model  
are investigated theoretically. 
It is found the long-time tail of
VACF disappears in a granular flow under the gravity. 

}

\begin{document}

\maketitle



\section{Introduction}


The long-time tails in the current autocorrelation functions are one of the important characteristics
in nonequilibrium physics \cite{alder,ernst71,Pomeau,Dorfman}.
It is known that the long tails are originated from the correlated collisions 
which cause anomalous transport behaviors such as the system size dependence of 
the transport coefficients \cite{Narayan,Murakami}.
The analysis of such behaviors is still an active subject in nonequilibrium physics. 

Although the long-time tail  $t^{-d/2}$ with the time $t$ and the spatial dimension $d$ 
even at equilibrium is well established, the interest in 
the long-tails in granular fluids is rapidly growing \cite{Kumaran,Orpe,Ahmad,Hayakawa}.
Most of the arguments focus on the velocity autocorrelation function (VACF) $C(t)$ which is defined by
\begin{equation}\label{VACF}
C(t)\equiv \lim_{V\to\infty}\frac{1}{dV}\sum_{i=1}^N\langle \bv{v}_i(0)\cdot\bv{v}_i(t)\rangle ,
\end{equation}
where 
$\bv{v}_i(t)$ is the velocity of $i-$th particle at time $t$, and $V$ is the volume of the system.
The average $\langle \cdots \rangle$ is taken over many different ensembles. 

So far theories and simulations to characterize granular fluids 
discussed in the case of freely-cooling case \cite{Ahmad,Hayakawa} 
or the flow under the simple shear.\cite{Otsuki07}
From these papers, we find that  (i) the long tail for VACF 
and the time-correlation of the shear stress exis but the time-correlation 
for the heat flux decays exponentially in freely cooling case\cite{Hayakawa}, and (ii) there is a cross-over 
from $t^{-d/2}$ to $t^{-(d+2)/2}$ for sheared granular gases.\cite{Otsuki07}
On the other hand, the existence of conventional long-time tails in VACF of the granular flow 
in a hopper has been reported experimently\cite{Orpe}, 
while there is no corresponding theoretical or numerical argument.

In this paper, thus,  we wish to clarify whether there exists the long tail of VACF  in the granular flow under the gravity.  
Our theoretical method is based on the classical one developed by Ernst et al. \cite{ernst71,Hayakawa}.
In the next section, we will introduce the model we use.  In $\S$ 3, we will present the details of the analysis.
In $\S$ 4, we will discuss and conclude our results.
In Appendix A, we show the details of the calculation of $C(t)$.

\section{Model} 

In this section, we introduce the model what we are analyzing. Although the experiment by Orpe and Kudrolli\cite{Orpe}
is designed for the dense flow in which the influence of the interstitial air is small, our theoretical tool
based on the method by Ernst {\it et al}.\cite{ernst71} might not be adequate for dense flows. On the other hand,
the drag from the side walls or the interstitial fluids plays a crucial role to achieve a steady granular flow.
Without the drag, particles are accelerated by the gravity and cannot reach a steady state.
Here, we introduce a fluid model to describe the granular flow under the gravity, which is a couple of equations
for the mass conservation and the momentum conservation:
\begin{eqnarray}
\partial_tn+\nabla\cdot(n\bv{u}) &=&0 , \label{mass} \\
\partial_t \bv{u}+\bv{u}\cdot\nabla\bv{u} &=& -\frac{T}{mn}\nabla n+\nu\nabla^2 \bv{u}+\frac{1}{\tau}[U(n)\bv{e}_z-\bv{u}],
\label{momentum}
\end{eqnarray}
where $n$ and $\bv{u}$ are the number density of grains and the coarse-grained velocity field, respectively. 
The parameters $T$, $m$, $\nu$ and $\tau$ repsectively represent the granular temperature, the mass of a particle, 
the effective kinetic viscosity and
the relaxation time to the optimal velocity $U(n)\bv{e}_z$ with the unit vector along $z-$axis. 
For granular particles immersed in fluids, $U(n)$ may be represented by\cite{HI} $U_1(1-\phi)^3/(1+2\phi+1.492\phi(1-\phi)^3)$
where $\phi$ is the volume fraction and $U_1=2a^2(\rho_p-\rho_f)g/9\eta$ is the one-particle sedimentation velocity 
with the radius of the particle $a$, the gravitational acceleration $g$, the density of the particle $\rho_p$, 
the density of the fluid $\rho_f$ and the fluid viscosity $\eta$. 
For traffic flows, we often use a phenomenological function for $U(n)$.\cite{bando}
In the model, we assume that the relaxation of the temperature is much faster than the other hyrodynamic variables.
This fast relaxation has been confirmed at least in the case of freely cooling granular gases\cite{Hayakawa}. 
 This model was original introduced 
as a one-dimensional traffic flow model by Karner and Konhauser\cite{KK}. 
Similar models have been used for the description
of the fluidized granular particles under the influence of air\cite{batchelor,sasa,komatsu}. 
Later, we have recognized that traffic flows and granular flows belong to the same universality class\cite{nishinomiya,wada}. 
It should be noted that the models for fluidized beds\cite{batchelor,sasa,komatsu} contain the pressure which depends
on the density, while our model adopts the equation of the state for the ideal gas. 
Although our model ignores the finite density effects in the pressure, the pressure includes
the most important contribution to keep a system uniform. 
On the other hand, the density dependence of the optimal velocity
$U(n)$ is essentially important to describe the emergence of the density waves.\cite{nishinomiya}
Therefore, we expect that the essence of hyrodynamic behaviors of granular flows is independent of the choice of the basic equation.

Let us call the model in eqs.(\ref{mass}) and (\ref{momentum}) the optimal velocity fluid model,
because the optimal velocity $U(n)\bv{e}_z$ plays important roles. 
We believe that the analysis of the optimal velocity model gives a prototype of granular flows and traffic flows.

\section{Theoretical analysis of VACF}

VACF is characterized by the fluctuations of hydrodynamic variables around the uniform flow
\begin{equation}\label{uniform}
n(\bv{r},t)=n_0, \quad \bv{u}(\bv{r},t)=U_0\bv{e}_z ,
\end{equation} 
where $U_0$ is the abbreviation of $U(n_0)$.
Ernst {\it et al}\cite{ernst71}. demonstrated that VACF can be approximated by
\begin{equation}\label{C(t)}
C(t)\simeq \int d\bv{v}_0 f_0(\bv{v}_0)\bv{v}_0\cdot 
\int \frac{d\bv{k}}{(2\pi)^d}\tilde{\bv{u}}_{\bv{k}}(t)\tilde{P}_{-\bv{k}}(t) ,
\end{equation}
where $\bv{v}_0$ is the relative one-particle velocity to the optimal velocity $U_0\bv{e}_z$,  
$f_0(\bv{v}_0)$ is the one-particle velocity distribution function, 
$\tilde{\bv{u}}_{\bv{k}}(t)$ and $\tilde{P}_{\bv{k}}(t)$ are respectively 
the Fourier transforms of $\bv{u}(\bv{r},t)-U_0\bv{e}_z$ and $P(\bv{r},t)$ 
with the probability density of a tracer particle $P(\bv{r},t)$.
From eq.(\ref{C(t)}), we can evaluate VACF once we know the behavior of $\tilde{\bv{u}}_{\bv{k}}(t)$ and 
$\tilde{P}_{\bv{k}}(t)$.

In general, it is not easy to solve the set of nonlinear equations (\ref{mass}) and (\ref{momentum}). 
When we restrict our interest, however, to the case of linearly stable situation around the uniform flow (\ref{uniform}),
we can simplify the argument. Such situations can be achieved for dilute flows or dense flows.
For simplicity, we only consider such linearly stable situations.

The probability density should obey the diffusion equation with the drift velocity: 
$(\partial_t+U_0\partial_z)P(\bv{r},t)=D\nabla^2P(\bv{r},t)$, where $D$ is the diffusion constant. 
Its solution is immediately obtained as
\begin{equation}\label{diffusion}
\tilde{P}_{\bv{k}}(t)=\tilde{P}_{\bv{k}}(0)\exp[-ik_zU_0 t-Dk^2t] .
\end{equation}
On the other hand, it is straightforward to obtain $d+1$ eigenvalues from the linearzed equations of (\ref{mass})
and (\ref{momentum}), which are $d-1$ degenerated $\sigma_{\perp}\equiv -i k_z U_0-\frac{1}{\tau}-\nu k^2$ and
\begin{eqnarray}\label{sigmapm}
\sigma_{\pm}&\equiv&\frac{1}{2}%
\left[
-\left(\frac{1}{\tau}+2ik_zU_0+\nu k^2\right)
\pm
\displaystyle\sqrt{\left(\frac{1}{\tau}+2ik_z U_0+\nu k^2\right)^2-
4\chi }
%
\right] 
\end{eqnarray}
with
\begin{equation}\label{chi}
\chi\equiv 
k^2\frac{T}{m}-k_z^2U_0^2-ik_z\left(\frac{U_0+U_0'n_0}{\tau}+U_0\nu k^2\right) ,
\end{equation}
where $U_0'\equiv \partial U/\partial n|_{n=n_0}$.
The most relevant eigenvalue to characterize the slow relaxation in the vicinity of $k\to 0$ is $\sigma_+$.
The uniform flow is no longer stable, when the real part of $\sigma_+$ can be positive if the condition $c_0^2>T/m$ 
is satisfied, where $c_0\equiv n_0U_0'$. 

For simplicity, however, we restrict our interest to the case of weakly stable case in which $T/m>c_0^2$ is
always satisfied in this paper.
Therefore, the approximate solutions of the linearized equations of (\ref{mass}) and (\ref{momentum}) are given by
\begin{eqnarray}\label{u_perp}
\tilde{\bv{u}}_{\bv{k}_{\perp}}(t)&\simeq&
\frac{e^{\sigma_{+}t}}{\sigma_+-\sigma_-}
[-\frac{T}{mn_0}i\bv{k}_{\perp}n_{\bv{k}}(0)
-\frac{T \bv{k}_{\perp}}{m(\sigma_+-\sigma_{\perp})}
(\bv{k}\cdot\tilde{\bv{u}}_{\bv{k}}(0)) \nonumber\\
& &+\left\{
\sigma_++ik_zU_0+\frac{k^2(T/m)+ik_z(n_0U_0'/\tau)}{\sigma_+-\sigma_{\perp}}
\right\}
\tilde{\bv{u}}_{\bv{k}_{\perp}}(0)
]
,
\\
\tilde{u}_{k_z}(t)&\simeq&
\frac{e^{\sigma_{+}t}}{\sigma_+-\sigma_-}
[
-\left(
\frac{ik_zT}{mn_0}-\frac{U_0'}{\tau}
\right)
\tilde{n}_{\bv{k}}(0)+
\frac{i n_0}{\sigma_+-\sigma_{\perp}}
\left(\frac{ik_zT}{mn_0}-\frac{U_0'}{\tau}\right)
(\bv{k}_{\perp}\cdot\tilde{\bv{u}}_{\bv{k}_{\perp}}(0))
\nonumber \\
& &+\left(\frac{k_{\perp}^2T}{m(\sigma_+-\sigma_{\perp})}+\sigma_++ik_z U_0\right)\tilde{u}_{k_z}(0)],
\label{u_z}
\end{eqnarray}
where $\bv{k}_{\perp}$ and $\tilde{\bv{u}}_{\bv{k}_{\perp}}$ are the wave vector in the horizontal plane and 
the velocity in the horizontal plane, respectively. Here, we stress that this solution is valid when $\exp(-t/\tau)$
is negligibly small. If $\tau$ is large enough, the relaxation term to the optimal velocity is negilible and 
the total energy of the system increases with time.

Substituting (\ref{diffusion}),(\ref{u_perp}) and (\ref{u_z}) into (\ref{C(t)}) we obtain
\begin{equation}\label{C=C_x+C_z}
C(t)=(d-1)C_x(t)+C_z(t),
\end{equation}
where 
\begin{eqnarray}\label{C_x}
C_x(t)&=&\frac{T}{dm}\int \frac{d\bv{k}}{(2\pi)^d}\frac{\exp[(\sigma_++ik_zU_0-Dk^2)t]}{\sigma_+-\sigma_-}
\nonumber\\
& & \times\left\{\frac{T(k^2-k_x^2)+ik_z (n_0U_0'/\tau)}{m(\sigma_+-\sigma_{\perp}}+\sigma_++ik_zU_0\right\},
\end{eqnarray}
and
\begin{equation}\label{C_z}
C_z(t)=\frac{T}{dm}\int\frac{d\bv{k}}{(2\pi)^d}\frac{\exp[(\sigma_++ik_zU_0-Dk^2)t]}{\sigma_+-\sigma_-}
\left\{\frac{k_{\perp}^2T}{m(\sigma_+-\sigma_{\perp})}+\sigma_++ik_zU_0\right\},
\end{equation}
where we use $\int d\bv{v}\bv{v}f_0(\bv{v})=0$, $\int d\bv{v} v_x^2f_0(\bv{v})=n T$, 
$\tilde{\bv{u}}_{\bv{k}}(0)\simeq (\bv{v}_0/n_0)$, and $\tilde{P}_{\bv{k}}(0)\simeq 1$.\cite{ernst71}
The evaluation of integrals in eqs.(\ref{C_x}) and (\ref{C_z}) is straightforward (The details of
 calculation for $d=3$ are presented in Appendix).
The long time behavior of $C_x(t)$ and $C_z(t)$ for $d=3$ are respectively given by
\begin{equation}\label{C_x2}
C_x(t)\simeq -\frac{\tau^2T^2}{4\pi^{3/2}m^2(D+\frac{T\tau}{m})\alpha^{1/2}}\frac{e^{-c_0^2t/(4\alpha)}}{t^{5/2}}
\propto  
-\frac{e^{-c_0^2t/(4\alpha)}}{t^{5/2}}
\end{equation}
and
\begin{equation}\label{C_z2}
C_z(t)\simeq -\frac{c_0^2\tau^2(T/m)(3c_0^2+\frac{2\alpha}{\tau}-\frac{T}{m})}{96\pi^{3/2}(D+\frac{T\tau}{m})\alpha^{5/2}}
\frac{e^{-c_0^2t/(4\alpha)}}{t^{3/2}}\propto
-\frac{e^{-c_0^2t/(4\alpha)}}{t^{3/2}},
\end{equation}
where $c_0\equiv n_0U_0'$ and $\alpha\equiv  D+\tau(T/m-c_0^2)$. From eqs.(\ref{C_x2}) and (\ref{C_z2}),
 it is obvious that $C_z(t)$ has the dominant contribution. Therefore, we obtain
\begin{equation}
C(t)\simeq C_z(t)\propto \frac{e^{-c_0^2t/(4\alpha)}}{t^{3/2}},
\end{equation}  
where the amplitude becomes negative when the system is in the weakly stable region of uniform flow.

It should be noted that we neglect the fast contribution $C_{\perp}(t)\propto \exp(-t/\tau)/t^{3/2}$ 
arising from $\sigma_{\perp}$, which corresponds to the correlation from the shear mode. This term becomes
relevant when the effects of the gravity and the relaxation to the optimal velocity is negligible to recover
the conventional tail $t^{-3/2}$ in the limit of $\tau\to \infty$. In this sense, our analysis presented here is valid
for $\tau<\alpha/c_0^2$.

Thus, the long-time behavior of VACF of the granular flow governed by eqs. (\ref{mass}) and (\ref{momentum}) has 
the exponential damping term as $C(t)\sim -e^{-c_0^2t/(4\alpha)}/t^{3/2}$ which is different from the case 
without the existence of systematic flows. 

\section{Discussion and Conclusion}

In this paper we have demonstrated the absence of the long-time tail in VACF for the granular fluid described by
(\ref{mass}) and (\ref{momentum}). Judging from the universality in the long-wave behaviors of granular flows and
traffic flows, our result is expected to be robust and applicable to any situation of flows of dissipative particles.
On the other hand, our result seems to be contradicted to the suggestion by Orpe and Kudrolli\cite{Orpe}.
Although we expect that the interpretation of their result may be consistent with ours if the flow speed is small enough,
we will need more systematic and careful treatments to clarify the relation between our analysis and their result.

In this paper, we only discussed the linearly stable case of the fluctuations around the uniform flow (\ref{uniform}).
However, one of the most important characteristics of the optimal velocity fluid model (\ref{mass}) and (\ref{momentum})
 exists in the emergence of the congestion or separation between congested region and sparse region. In this sense,
our analysis is far from completion. 
It is known that there is the frequency spectrum obeying $f^{-4/3}$ for granular flows\cite{nishinomiya,moriyama,hisao05},
the relation between the present analysis and the previous ones is not clear. We will have to clarify the relation in
future.

In conclusion, VACF in the granular flow governed by the optimal velocity fluid model (\ref{mass}) and (\ref{momentum}) 
does not have the long-time tail but has the exponential decay. The amplitude of the exponential function becomes
zero when there is no optimal speed $U_0$. Our result strongly suggests that we need careful interpretation of
the experiment by Orpe and Kudrolli\cite{Orpe}.

This work is partially supported by Ministry of Education, Culture, Science and Technology (MEXT), Japan
 (Grant No. 18540371).

\appendix
\section{The calculations of $C_x(t)$ and $C_z(t)$}

In this appendix, we explicitly calculate (\ref{C_x}) and (\ref{C_z}).
For simplicity, we restrict our interest to 3-dimensional cases.

For the calculation of $C_x(t)$ and $C_z(t)$ we need to know the expressions of $\sigma_+$, $(\sigma_+-\sigma_-)~{-1}$
and $\{(\sigma_+-\sigma_-)(\sigma_+-\sigma_{\perp})\}^{-1}$ in the long-wave expansion as
\begin{eqnarray}\label{a1}
\sigma_+&\simeq& -i(c_0+U_0)\tau^2k_z+\tau\left(c_0^2-\frac{T}{m}\right)k_z^2-\frac{T\tau}{m}k_{\perp}^2+\cdots \\
\label{a2}
(\sigma_+-\sigma_-)^{-1}&\simeq&
\tau+2i \tau^2 c_0 k_z+\tau^2\left(2\tau\left(\frac{T}{m}-3c_0^2\right)-\nu\right)k_z^2 \nonumber \\
& &+\tau^2\left(\frac{2T\tau}{m}-\nu\right)k_{\perp}^2+\cdots \\
\frac{1}{(\sigma_+-\sigma_-)(\sigma_+-\sigma_{\perp})}
&\simeq&
\tau^2+3ic_0\tau^3k_z+\tau^3\left(\tau\left(3\frac{T}{m}-10c_0^2\right)-2\nu\right)k_z^2 \nonumber\\
& &+
\tau^3\left(\frac{3T\tau}{m}-2\nu\right)k_{\perp}^2+\cdots.
\label{a3}
\end{eqnarray}

Let us calculate $C_x(t)$ at first.
Substituting (\ref{a1})-(\ref{a3}) into (\ref{C_z}) with some manupulation we obtain
\begin{equation}\label{a4}
C_x(t)=-\frac{T^2\tau^2}{24\pi^3m^2}\int d\bv{k} k_x^2
\exp\left[-i c_0k_zt-\alpha k_z^2t-\left(D+\frac{T\tau}{m}\right)k_{\perp}^2t\right]
\end{equation}
Introducing the cylidrical coordinate $k_x=k_{\perp}\cos\theta$ with 
$\int d\bv{k}=\int_0^{\infty}dk_{\perp}\int_0^{2\pi}d\theta\int_{-\infty}^{\infty}dz$ we obtain (\ref{C_x2}), where
we use $\int_{-\infty}^{\infty}dk_z\exp[-(ic_0k_z+\alpha k_z^2)t]=\sqrt{\pi/(\alpha t)}e^{-c_0^2t/(4\alpha)}$
and $\int_0^{\infty}dk k^3\exp[-\beta k^2t]=(2\beta t)^{-1}$.
 
Next, we calculate $C_z(t)$. Substituting (\ref{a1})-(\ref{a3}) into (\ref{C_z}) we obtain
\begin{equation}\label{a5}
C_z(t)=\frac{T}{24\pi^3m}\left[-c_0\tau C_z^{(1)}(t)+\tau^2\left(3c_0^2-\frac{T}{m}\right)C_z^{(2)}(t)\right],
\end{equation}
where
\begin{eqnarray}\label{a6}
C_z^{(1)}(t)&\equiv& i \int d\bv{k}k_z 
\exp\left[-\left(ic_0k_zt+\alpha k_z^2+\left(D+\frac{T\tau}{m}\right)k_{\perp}^2\right)t\right] \nonumber\\
&=&\frac{\pi^{3/2}c_0}{2\alpha^{3/2}\beta t^{3/2}}\exp\left[-\frac{c_0^2t}{4\alpha}\right], \\
C_z^{(2)}(t)&\equiv&
\int d\bv{k}k_z^2 
\exp\left[-\left(ic_0k_zt+\alpha k_z^2+\left(D+\frac{T\tau}{m}\right)k_{\perp}^2\right)t\right] \nonumber\\
&=&-\frac{\pi^{3/2}c_0^2(1-\frac{2\alpha}{c_0^2t}) }
{4(D+\frac{T\tau}{m})\alpha^{5/2}t^{3/2} }
\exp\left[-\frac{c_0^2t}{4\alpha}\right]
\label{a7}
\end{eqnarray}
where we use $\int_0^{\infty}dz z^2 \cos(c_0zt)e^{-\alpha z^2 t}=
\frac{\sqrt{\pi}(2\alpha-c_0^2t)}{4\alpha^{3/2}t^{3/2}}e^{-c_0^2t/(4\alpha)}$. 
Substituting (\ref{a6}) and (\ref{a7}) into (\ref{a5}) we obtain (\ref{C_z2}) as the long time behavior.


\begin{thebibliography}{99}
\bibitem{alder} B. J. Alder and T. E. Wainwright, Phys. Rev. Lett. {\bf 18} (1997) 988, 
Phys. Rev. A {\bf 1}(1970) 18.
\bibitem{ernst71}   M. H. Ernst, E. H. Hauge and J. M. J. van Leeuwen, Phys. Rev. A {\bf 4} (1971) 2055.


\bibitem{Pomeau} Y. Pomeau and P. R{\'e}sibois, Phys. Rep. {\bf 19} (1975) 63.
\bibitem{Dorfman}  J. R. Dorfman, T. R. Kirkpatrick, and J. V. Sengers,
Ann. Rev. Phys. Chem.  {\bf 45} (1994) 213.

\bibitem{Narayan}  O. Narayan  and S. Ramaswamy, Phys. Rev. Lett {\bf 89} (2002) 200601.

\bibitem{Murakami} T. Murakami, T. Shimada, S. Yukawa and N. Ito, J. Phys. Soc. Jpn. {\bf 72} (2003) 1049.

\bibitem{Kumaran} V. Kumaran, Phys. Rev. Lett. {\bf 96} (2006) 258002.
\bibitem{Orpe}  A. V. Orpe and A. Kudrolli 
Phys. Rev. Lett. {\bf 98} (2007) 238001.

\bibitem{Ahmad} S. R. Ahmad and S. Puri, Phys. Rev. E {\bf 75} (2007), 031302.

\bibitem{Hayakawa} H. Hayakawa and M. Otsuki, Phys. Rev. E {\bf 76} (2007), 051304.
\bibitem{Otsuki07} M. Otsuki and H. Hayakawa, arXiv:0711.1421
\bibitem{HI} H. Hayakawa and K. Ichiki, Phys. Rev. E {\bf 51} (1995), R3815.
\bibitem{bando} M. Bando, K. Hasebe, A. Nakayama, A. Shibata and A. Sugiyama, 
Phys. Rev. E {\bf 51} (1995), 1035.
\bibitem{KK} B. S. Kerner and P. Konhauser, Phys. Rev. E {\bf 48} (1993) 2335.
\bibitem{batchelor} G. K. Batchelor, J. Fluid Mech. {\bf 193} (1988), 75.
\bibitem{sasa} S. Sasa and H. Hayakawa, Europhys.Lett. {\bf 17} (1992) 685.
\bibitem{komatsu} T. S. Komatsu and H. Hayakawa, Phys. Lett. A {\bf 183} (1993) 56.
\bibitem{nishinomiya} H. Hayakawa and K. Nakanishi, Prog. Theor. Phys. Suppl. {\bf 130} (1998) 57.
\bibitem{wada} S. Wada and H. Hayakawa, J. Phys. Soc. Jpn. {\bf 67} (1998) 763.
\bibitem{moriyama} O. Moriyama, N. Kuroiwa and M. Matsushita and H. Hayakawa, Phys. Rev. Lett. {\bf 80} (1998) 2833.
\bibitem{hisao05} H. Hayakawa, Phys. Rev. E {\bf 72} (2005) 031102.

\end{thebibliography}
\end{document}